\documentclass[times,12pt]{article}
\date{}

\begin{document}

\title{{\LARGE\sf } The Effect of Pure State Structure on Nonequilibrium Dynamics}
\author{
{\bf C.M. Newman}\\
{\small \tt newman\,@\,cims.nyu.edu}\\
{\small \sl Courant Institute of Mathematical Sciences}\\
{\small \sl New York University}\\
{\small \sl New York, NY 10012, USA}
\and
{\bf D.L. Stein}\\
{\small \tt daniel.stein\,@\,nyu.edu}\\
{\small \sl Dept.\ of Physics and Courant Institute of Mathematical Sciences}\\
{\small \sl New York University}\\
{\small \sl New York, NY 10012, USA}
}

\maketitle

\begin{abstract}
Motivated by short-range Ising spin glasses, we review some rigorous
results and their consequences for the relation between the number/nature
of equilibrium pure states and nonequilibrium dynamics. Two of the
consequences for spin glass dynamics following a deep quench to a
temperature with broken spin flip symmetry are: (1) Almost all initial
configurations lie on the boundary between the basins of attraction of {\it
multiple\/} pure states. (2) Unless there are uncountably many pure states
with almost all pairs having zero overlap, there can be no equilibration to
a pure state as time $t \to \infty$.  We discuss the relevance of these
results to the difficulty of equilibration of spin glasses.  We also review
some results concerning the ``nature vs.~nurture'' problem of whether the
large-$t$ behavior of both ferromagnets and spin glasses following a deep
quench is determined more by the initial configuration or by the dynamics
realization.
\end{abstract}

{\bf KEY WORDS:\/}  spin glass; nonequilibrium dynamics; deep quench;
stochastic Ising model; broken ergodicity; coarsening;
persistence; damage spreading.

\small
%\vskip 1.25in
%$^*$ A wholly owned subsidiary of Emirate Investments, Ltd.
\renewcommand{\baselinestretch}{1.25}
\normalsize
\newpage
\section{Introduction}
\label{sec:intro}

Experimental signatures of laboratory spin glasses --- irreversibility,
history dependence, aging --- demonstrate that these systems are out of
equilibrium during the timescale of most, if not all, experimental
measurements~\cite{BY86,BCKM98,aging1,aging2,aging3,aging4,aging5,aging6,aging7,aging8,aging9,aging10,LOHOV88,HLOOV92,JOH96,VHOBC97,JVHPN98,RMB00,RTZ00,VDAHB00}.
This puts us in the unusual position of attempting to explain the
nonequilibrium dynamics of a system whose equilibrium statistical mechanics
have yet to be worked out, or even understood on the most basic qualitative
level.  We still don't know, for example, whether there exists a true
equilibrium phase transition to a spin glass phase in {\it any\/} dimension
in the usual short-range models that are believed to describe laboratory
spin glasses (many of which aren't short-range at all.  This leads to the
question of how well short-range models do in fact represent them --- but
since none of the models are at all undersood, that question remains moot
for now.)  Supposing that there is a phase transition in short-range
models, we don't know their low-temperature properties, such as pure state
multiplicity and structure.  We don't even know what their zero-temperature
(i.e., ground state) properties look like.  (It would be an overstatement,
however, to say we don't know anything.  Numerical work has provided some
valuable
insight~\cite{Ogielski85,OM85,RBY90,KY96,PY99a,PY99b,Middleton99,Middleton00,Hartmann00,KM00,PY00,HKM00,MP00,MP01,HYD04},
there exist several competing pictures for the spin glass
phase~\cite{MPV87,MPRRZ00,Mac84,BM85,FH86,FH88a,FH88b,BM87,NS97,NS98}, and
rigorous and nonrigorous work of the authors has effectively ruled out some
scenarios~\cite{NS03b}.  But we clearly still have a long way to go.)

This lack of progress is partly responsible for the viewpoint advocated by
some that the only physics of spin glasses really worth looking at is their
nonequilibrium dynamics.  Although this seems to us a bit premature given
that we don't really know what the equilibrium properties look like,
there's no question that even if we did have a better picture of the
equilibrium thermodynamics, we could still be a long way from explaining
the many experimental observations of spin glass behavior.

Related to this, but not quite the same, is the prevalent viewpoint that
the nonequilibrium dynamics of spin glasses (or any system with many
competing thermodynamic phases) is sharply separated from their equilibrium
behavior, in particular, their possessing many pure states.  The actual
presence of many pure states may or may not exist in real spin glasses, but
the possibility has created substantial excitement about these systems and
spurred numerous theoretical and numerical investigations.  The basic idea
behind this viewpoint lies in the dynamical invariance of pure states. That
is, dynamics takes a spin configuration belonging to a pure state to
another in the same pure state, but never to a configuration in a different
pure state; equivalently, pure states are separated by dynamically
insurmountable barriers. It therefore seems initially reasonable to
conclude that the number of equilibrium pure states is dynamically
irrelevant: since dynamics always occurs within a single pure state, no
matter how long the timescale, knowledge of the equilibrium pure state
structure doesn't tell one very much about the nonequilibrium dynamics.

The main point of this paper is to convince the reader that this idea is
incorrect, and that the two (equilibrium structure and nonequilibrium
dynamics) are in fact very much interrelated.  Of course, the context in
which an experiment is done is crucial in any statement of this kind. To
support the above claim, we will review earlier papers of the authors, and
discuss some mathematically rigorous theorems there that demonstrate a deep
interconnection between equilibrium pure state structure and nonequilibrium
dynamics in one of the farthest-from-equilibrium situations that has been
studied: a sudden quench from very high to low temperature.  As we proceed,
we'll show along the way that a number of other widely-held beliefs break
down in this much studied situation.  One of these is the common
assumption that the union of the basins of attraction of all of the pure
states fills up most of the available state space, and that the boundaries
between the pure states are lower dimensional and thereby form a set of
measure zero in the set of all spin configurations (in fact, the opposite
will turn out to be true).  Another is that if many pure states are present
in an infinite system, time averages don't agree with Boltzmann averages
(in principle and perhaps even in practice, they can agree).

Our approach is general, and covers both ordered and disordered, Ising and
non-Ising systems, although we will usually focus our attention on
nearest-neighbor Ising spin glasses for specificity. We make no {\it a
priori\/} assumptions about the real-space or state-space structure of the
low-temperature spin glass phase, but instead derive several general
principles and then explore their consequences.

We emphasize that our discussion centers on pure states, not metastable
states.  Crudely put, metastable states are surrounded by barriers that
remain of $O(1)$ irrespective of the size of the system, while pure states
are surrounded by barriers that diverge in the thermodynamic limit.  (More
precise definitions can be found in~\cite{NS03b}, but in this paper we
sacrifice some mathematical precision for readability.)  Metastable states
are often proposed as responsible for the anomalous dynamical behavior of
spin glasses.  While we have no argument with this, we question the
usefulness of the usual practice of inserting metastability by hand,
requiring a guess as to the structure (usually in state space) and nature
of the metastable states.  (In fact, much of this structure can be
determined {\it ab initio\/} --- e.g., a rigorous discussion of
metastability in spin glasses and disordered systems can be found
in~\cite{NS99c}.)

To keep the paper   reader-friendly, 
we will present theorems without their
proofs, which can be found in the cited references.  

\section{Pure states, dynamics, and equilibration}
\label{sec:measures}

Although the preceding discussion uses familiar terms and notions, their
actual meanings require some work to pin down.  For example, what does
equilibration following a deep quench mean in an infinite system?  Can it
occur on any finite timescale?  What does it mean for a system to evolve,
or settle into, or even to ``spend all its time inside,'' a single pure
state?  In order to proceed, we need to clarify these notions.

For specificity,we will mostly, but not exclusively, consider the
Edwards-Anderson (EA) Ising spin glass~\cite{EA75} in zero external
magnetic field.  Its Hamiltonian is given by:
\begin{equation}
\label{eq:EA} 
{\cal H}_{\cal J} = -\sum_{<xy>} J_{xy}\sigma_x \sigma_y\, ,      
\end{equation}
where the sites $x,y\in {\bf Z}^d$ and the 
sum is taken over nearest neighbors only.  The 
couplings $J_{xy}$ are independent random variables,
whose common probability density is 
symmetric about zero; we let ${\cal J}$
denote a particular realization of the couplings.

We next need to specify the dynamics.  We are interested in the
experimental situation in which a spin glass dynamically evolves following
a deep quench. We model this as the quench of an infinite system governed
by the Hamiltonian~(\ref{eq:EA}) from infinite to low temperature.  This is
done by first choosing the initial   (time $t=0$) spin configuration
$\sigma^0$ from the infinite-temperature ensemble where the individual
spins are independent random variables equally likely to be $+1$ or
$-1$. We then use the usual Glauber dynamics, which is easiest to describe
at zero temperature; there, flips that are energy lowering occur with
probability $1$, flips that are energy neutral (neither lowering nor rasing
the energy) occur with probability $1/2$, and flips that are energy raising
occur with probability $0$.  The exact choice of spin flip rates plays no
role in our analysis as long as detailed balance is satisfied, so we may
take the rate of considering flips to be $1$.  At nonzero temperature any
dynamical rule consistent with detailed balance can be used, such as the
usual Metropolis or heat bath dynamics.

At  
 any temperature $T$, 
we denote by $\omega$ a given realization of the
dynamics (i.e., of the order in which spins are chosen to determine whether
they flip according to the dynamical rules, combined with the outcome of
each of these trials). Any $\omega$ can be regarded as a collection of
random times $(t_{x,i}:\,x \in {\bf Z}^d,\,i=1,2,\dots)$, each specifying
when a spin flip at site $x$ is considered 
(forming a Poisson process   in time for
each $x$) along with random numbers $u_{x,i}$ that determine if the flip is
taken.

So there are three sources of randomness in the problem: the couplings, the
initial spin configuration, and the dynamics.  Specific realizations for
any given dynamical run are denoted respectively by ${\cal J}$, $\sigma^0$,
and $\omega$.  All three are needed to determine $\sigma^t$, the spin
configuration at time $t$. We always take ${\cal J}$ to be fixed during any
single run, to correspond to experimental situations on laboratory
timescales.

\subsection{Equilibrium states}
\label{subsec:es}

Consider now a finite volume $\Lambda_L$, say a cube, of linear size $L$
(which may be arbitrarily large) centered at the origin.  For a given
boundary condition (b.c.), the equilibrium finite-volume Boltzmann
distribution is given at temperature $T$ by
\begin{equation}
\label{eq:Boltzmann}
\rho_{{\cal J},T}^{(L)}(\sigma)=Z_{{\cal J},L,T}^{-1} \exp \{-{\cal H}_{{\cal J},L}(\sigma)/k_BT\}\quad ,
\end{equation}
where the finite-volume spin configurations $\sigma$ are restricted to
those obeying the b.c.~and the partition function $Z_{{\cal J},L,T}$ is
such that the sum of $\rho_{{\cal J},T}^{(L)}$ over all spin configurations
in $\Lambda_L$ yields one.

The quantity $\rho_{{\cal J},T}^{(L)}(\sigma)$ is of course simply a
probability measure: it describes at fixed $T$ the probability of a given
spin configuration $\sigma^{(L)}$ obeying the specified boundary condition
appearing within $\Lambda_L$.  One is also interested in infinite-volume
measures $\rho = \rho_{{\cal J},T}(\sigma)$ that specify the probability of
{\it any\/} spin configuration appearing within $\Lambda_{L'}$, for every
$L'$; we call such a $\rho$ a Gibbs state.  The infinite-volume limit of
any convergent sequence of $\rho_{{\cal J},T}^{(L)}(\sigma)$'s with any set
of b.c.'s (which can vary with $L$) specifies a Gibbs state; one or many
Gibbs states may exist, depending on the system, temperature, and
dimension.  (The assertion sometimes made that such infinite-volume
quantities may not exist or make sense for spin glasses is incorrect: it is
easily shown that they do exist and that they govern the equilibrium
behavior of spin glasses; we just don't happen to know what they look
like.)  Gibbs states may be pure or mixed; see Sect.~4.1 of~\cite{NS03b}
for definitions and a detailed discussion.  (The term ``pure state'' used
throughout this paper refers to these pure equilibrium states.)

\subsection{Dynamical states}
\label{subsec:ds}

To discuss the questions asked at the beginning of this section in a
meaningful way, we also need to specify a {\it dynamical\/} probability
measure with which Gibbs states can be compared.  Consider then the
infinite-volume spin configuration $\sigma^t$ at time $t>0$.  As already
mentioned, $\sigma^t$, a dynamical Markov process, depends on ${\cal J}$,
$\sigma^0$, and $\omega$, but for ease of notation this dependence will be
suppressed.  We define the dynamical probability measure $\nu_{t}(\sigma)$
as the distribution of $\sigma^t$ over the dynamics $\omega$ for fixed
${\cal J}$ and $\sigma^0$.  That is, $\nu_t$ tells us the probability for
each $L$ of finding a particular spin configuration $\sigma^{(L)}$ within
$\Lambda_L$ at time $t$ for a given ${\cal J}$ and starting configuration
$\sigma^0$.  We will also sometimes consider a measure $\nu_{t,\tau}$ which
is the distribution of $\sigma^t$ over that part of the dynamics between
times $t-\tau$ and $t$, so that $\nu_{t,\tau}$ depends on ${\cal J}$,
$\sigma^0$ and $\omega$ (before time $t-\tau$).

Because detailed balance is satisfied by the dynamics, we expect that after
a sufficiently long time $t$, the probability assigned by the dynamical
measure $\nu_t(\sigma)$ (or by $\nu_{t,\tau}$ with both $t$ and $\tau$
large) to a given spin configuration $\sigma^{(L)}$ within $\Lambda_L$ will
approach that assigned to that same configuration by some Gibbs state
$\rho_{{\cal J},T}$, and that this will be true for any $L$ and
$\sigma^{(L)}$ (of course, how long one has to wait before this occurs will
depend on $L$).  Although this may be surprising at first, especially in
light of assertions that equilibrium states are of little relevance for the
nonequilibrium dynamics of infinite systems, it is to be expected if the
common conjecture holds for Glauber dynamics (even in infinite volume) that
at positive temperature {\it only\/} Gibbs states (where the probability of
appearance of spin configurations are given by the Boltzmann distribution)
are stationary.  Note that the $\rho_{{\cal J},T}$ in this discussion need
not be a pure state.

\subsection{Equilibration and nonequilibration}
\label{subsec:equil}

Now we can address the question of what it means for the system to evolve
into (or within) a specific {\it pure\/} state.  Since this involves some
sort of equilibration, we need to address first the broader question of
what equilibration means in an infinite system.  This has been subject to
various interpretations.  A common viewpoint~\cite{Bray94} is that infinite
systems (following a deep quench, say) never reach equilibrium in any
finite time; for example, in the homogeneous ferromagnet domains of
positive and negative magnetization increase with time but are never
infinite on any finite timescale.  This is of course true, but we do not
find it to be a useful way of looking at equilibration considering that
equilibrium states are really a {\it local\/} concept (see,
e.g.,~\cite{NS06a}). %MAYBE THEY'RE A GLOCAL CONCEPT?

Instead, we propose the following~\cite{NS99a}: if for any finite region
$\Lambda_L$ there exists a time
  $t^*_L<\infty$ after which the
distribution %$\nu_(\tau)(\sigma)$ 
$\nu_t$ (or $\nu_{t,\tau}$), restricted to $\Lambda_L$, is 
(approximately) the same as some %$\rho_{{\cal J},T}$,
{\it pure\/} state $\alpha$, then we say that
the system has equilibrated in finite time.  (It doesn't matter that the
equilibration time depends on the region size.) So, in the case of the
ferromagnet, our definition implies that an infinite ferromagnet
equilibrates if, for any region of any size, domain walls (between positive
and negative magnetization) cease to move across the region after some
finite time (depending on the region).

Is this definition trivial?  No, because it may be that such ``local
equilibration'' does {\it not\/} occur.  As a specific example, it doesn't
  occur
for $2D$ ferromagnets at low $T$; 
it can be rigorously proved that, following a deep
quench, for any finite region, domain walls continually sweep across it
(presumably at increasingly widely spaced time intervals) for all
time~\cite{NNS00}.  (We don't know 
  for sure what happens above two dimensions, but
we conjecture, based on numerical work of Stauffer~\cite{Stauffer94}, that
this ``local nonequilibration'' holds for the homogeneous ferromagnet up to
four dimensions while at five dimensions and above local equilibration occurs.)

We will discuss local nonequilibration and its consequences more in
Sec.~\ref{sec:effect}.  For now, however, we return to the question of pure
states.  If local equilibration has occurred and the dynamical distribution
$\nu_t$ (or $\nu_{t,\tau}$) has approached a pure state $\alpha$, then the
system has settled into that pure state.  In that case the entire
(infinite) system settles into $\alpha$; it cannot be that different
regions have settled into different pure states (e.g., the positive and
negative magnetization states in the homogeneous ferromagnet).  But if
local equilibration does {\it not\/} occur, it should still be true that
(after some finite time depending on the region) any finite region will
approximate a pure state (in the sense described above) at most times (the
exceptional times being when a domain wall between different pure states
sweeps across the region).  The details behind this assertion can be found
in Sec.~2 of~\cite{NS99a}.

In the case of local equilibration, the pure state $\alpha$ that the region
$\Lambda_L$ has settled into at time $t$ should depend on ${\cal J}$,
$\sigma^0$, and $\omega$.  It is for that reason the the theorem discussed
in the next section comes as a surprise, and has some far-reaching
consequences.

\section{Basins of attraction of pure states form a set of measure zero}
\label{sec:boa}

We have seen that the problem of equilibration at positive temperature
boils down to the question of whether, on every fixed (and arbitrarily
large) lengthscale $L$ the dynamical measure $\nu_t$ (more precisely
$\nu_{t,\tau}$ for some $\tau(t) \to \infty$ with $t$) settles down to a
pure state Gibbs measure $\rho^\alpha$ that is independent of time (after a
timescale depending on $L$).  Even if this doesn't occur, at {\it most\/}
times larger than some $t^*_L$, the dynamical measure $\nu_t$ (more
precisely $\nu_{t,\tau(t)}$) will approximate a pure state $\alpha(t)$; in
this case, at presumably widely spaced times a domain wall (here to be
thought of as the boundary separating two distinct pure states) sweeps
across $\Lambda_L$, changing the pure state seen within the volume.  Even
so, if one simply chooses a fixed, arbitrary time much larger than $t^*_L$,
with high probability the finite volume $\Lambda_L$ will be found inside
some thermodynamic pure state.

This naturally raises the question: how does the system evolve, and what
determines its long-time dynamical evolution?  A standard viewpoint is that
the system should evolve into some pure state, depending on initial
conditions, within which it remains forever after.  How might this happen?
Perhaps some part of the system finds a pure state before other parts
(because it was fortuitously close to one to begin with), and this region
then grows so that eventually any part of the system can be found in that
same pure state.  If other parts of the system fall into different pure
states, then upon their boundaries meeting one such state, perhaps the one
in the larger volume, would presumably ``win''.

Of course, if there is only one pure state governing the equilibrium
thermodynamics (such as, e.g., the paramagnetic state in the uniform
ferromagnet above $T_c$), then  
a simple version of something like this happens, and the
system does settle into that pure state.  However, if there is more than
one pure state, whether two, ten, or an infinite number, then the scenario
described above {\it never\/} happens  
for $\nu_t$.  This claim follows from the
following theorem:

{\it Theorem 1\/}~\cite{NS99a}: Given some ${\cal J}$ and $T>0$, assume
that for almost every $\sigma^0$, $\nu_t$ converges to a limiting {\it
pure\/} Gibbs state $\nu_{\infty}$ as $t\to\infty$.  Then $\nu_{\infty}$ is
the same pure state for almost every $\sigma^0$.

The proof of Theorem~1 is fairly short and appears in~\cite{NS99a}.  Here we
are concerned only with its consequences.  The theorem applies regardless
of whether the system's equilibrium thermodynamics is governed by a single
pure state (in which case the conclusion is trivial), or many. 

In the latter case, of course we expect that the dynamical outcome {\it
should\/} depend on the starting configuration.  There are three possible
ways to resolve this apparent contradiction:

1) $\nu_{\infty}$ is not a Gibbs state.  This would violate the expected
$T>0$ behavior discussed in Section~\ref{sec:measures} and hence we
  discard this possibility.

2) $\nu_{\infty}$ is a {\it mixed\/} Gibbs 
state (which may or may not depend on $\sigma^0$).

3) $\nu_t$ does {\it not\/} converge  
(to a single limit) as $t \to \infty$.  

Let's consider the second and third possibilities in more detail.
Possibility (2) implies that at (suitably large) time $t$, the dynamical
measure $\nu_{t,\tau(t)}$ is approximately a pure state $\rho^{\alpha(t)}$,
but that pure state depends not only on $\sigma^0$ (as expected) but also
on $\omega$ (i.e., the dynamical realization between times 0 and
$t-\tau$). So, while the system might ``land'' in a pure state in the sense
that $\alpha(t)$ converges to some $\alpha(\sigma^0,\omega)$, the limiting
pure state is almost never determined solely by $\sigma^0$.

But this has a strong consequence for the geometry of the state space
structure.  The basin of attraction of a pure state $\bar{\alpha}$ is the
set of configurations $\sigma^0$ such that $\alpha(\sigma^0,\omega) =
\bar{\alpha}$ for almost every $\omega$ (see Ref.~\cite{vEvH84} for related
discussions).  Therefore, if many pure states exist (and the conclusion of
Theorem~1 is not valid), then {\it the union of all their basins of
attraction must form a set of measure zero in the space of $\sigma^0$'s\/};
i.e., the configuration space resulting from a deep quench is all
``boundary'' in the sense that almost every initial configuration could
land in one of several (or many) pure states depending on the realization
of the dynamics (if it lands at all).

This result is perhaps counterintuitive.  In the Introduction, it was noted
that the context in which an experiment is done is crucial to the
interpretation of statements relating equilibrium thermodynamic structure
to nonequilibrium dynamical behavior.  This example provides an important
illustration of this. Because the quench is from a very high (formally,
infinite) to a low temperature, the relevant configuration space that the
system must explore (at least for small time $t$) is effectively the one
prevailing at high temperatures.  However, the pure state basins that the
system evolves to are those relevant to the (low) temperature that
determines the dynamical rules.  Looked at in this way, it may not seem
quite so strange (in fact, it seems quite natural) that the pure states
form a set of measure zero in the configuration space.  But then this also
illustrates that any statements relating or contrasting equilibrium
thermodynamics with nonequilibrium dynamics cannot in general be made
independently of the dynamical process under consideration.

We note finally that Theorem~1 may be relevant to damage
spreading~\cite{Bag96,Grass95,JR94}, where one asks whether the damage
(i.e., discrepancy) between $\sigma^t$ and $\sigma'^{\,t}$ (with a single
$\omega$) grows as $t \to \infty$. Theorem~1 suggests that if damage
spreading occurs, then $\nu_t$ doesn't converge to a single pure state
(e.g., it might converge to a mixed state, as above).

\section{Effect of pure states on nonequilibrium dynamics}
\label{sec:effect}

Before discussing possibility~(3), let us consider the physical picture
implied by Theorem~1.  Roughly speaking, some time after an initial quench
the system will form domains, whose average size increases with time,
corresponding to the different pure states.  This scenario has been
analyzed for the two-state droplet picture~\cite{FH88a,KH88,TH96}.  It is also
a well-known scenario for coarsening in a ferromagnet following a deep
quench~\cite{Bray94}. (Of course, in contrast to the spin glass case,
one {\it does\/} know how to prepare a ferromagnet in a pure state; for a
general discussion, see Ref.~\cite{Palmer82}.)

But if possibility~(3) holds, then even after a region has settled into a
pure state, it remains ``alive'' dynamically: the pure state in that region
will eventually change.  In Sec.~\ref{subsec:equil}, we defined system
equilibration after a finite time in terms of {\it local\/} equilibration:
for any region $\Lambda_L$, there exists a time $t_L$ after which domain
walls cease to move across the region.  Possibility~(3) would then
correspond~\cite{NS99a} to local {\it non-equilibration\/} (LNE) on any
finite lengthscale. (We note that Possibility~(2) could occur either with
local equilibration or non-equilibration.)

So LNE means that in any fixed finite region, the system never settles down
into a pure state.  Domain walls do not simply move farther from the region
as time progresses, but continually return and sweep across it, changing
the state within. If LNE occurs in the spin glass, it would force us to
revise the usual dynamical definition~\cite{BY86} of the EA order
parameter.  It could also mean that, for infinite systems, time averages
and Gibbs averages could agree, despite the presence of many pure states.

We will return to this in Sec.~\ref{sec:fixed} after investigating LNE in
more detail by means of the next theorem, which applies to both homogeneous
and disordered systems and ties together equilibrium pure state structure
with nonequilibrium dynamics.

\medskip

{\it Theorem 2.\/}~\cite{NS99a} Let ${\cal N}$ be the number of pure states
in the EA model for fixed $T$ and $d$, and suppose that $T$ and $d$ are
such that all pure states have nonzero EA order parameter (implying that
${\cal N}$, which is the same for almost every ${\cal J}$, cannot be~$1$).
If ${\cal N}$ is countable (including a countable infinity), then LNE
occurs. 

An immediate consequence of Theorem~2 is that if LNE does not occur (and
the limiting pure states have nonzero $q_{EA}$), then there must be an
uncountable number of pure states.  Furthermore, the proof of Theorem~2
(see~\cite{NS99a}), which is based on overlaps, shows that almost every
pair of these pure states has overlap equal to zero.  This shows that, as
claimed in the Introduction, nonequilibrium dynamics can provide important
information on the structure of equilibrium pure states, and vice-versa.

It also suggests a dynamical test of the two-state picture: search for LNE
in the dynamical measure $\nu_t$ or $\nu_{t,\tau}$.  If LNE does not occur,
then the two-state picture has been ruled out --- there must be an
uncountable number of pure states with almost all pairs having overlap zero
(consistent with the results of Ref.~\cite{NS97}).  If LNE does occur then
neither the two-state nor the many-state pictures have been ruled out.  (It
is not very clear how one might go about observing LNE in a spin glass,
where, unlike the ferromagnet, one doesn't know what a domain wall looks
like.  For a discussion of how this might be accomplished,
see~\cite{NS99a}.)

Theorem 2 also implies that LNE occurs at small positive temperature in the
$2D$ uniform Ising ferromagnet and the random Ising ferromagnet for $d<5$.
In the former case this result was extended to zero temperature (using
different arguments) in~\cite{NNS00}.  It was also shown there that for
many systems (e.g., spin glasses and random ferromagnets where the common
distribution of the $J_{xy}$'s is continuous) that $\sigma^t$ {\it does\/}
converge to some limit at $T=0$.  (There are also systems, such as the $\pm
J$ spin glass on the square lattice at $T=0$, where some spins flip only
finitely many times and some spins flip infinitely often~\cite{GNS00}.)  In
light of these results, we restrict the term LNE to $T>0$, since in the
zero-temperature situations where $\sigma^t$ converges, the limit
configuration is typically only metastable rather than a ground state and
so equilibration has not really occurred.  In these systems one can define
a dynamical order parameter, related to the autocorrelation, that does {\it
not\/} decay to zero.

\section{More about local non-equilibration}
\label{sec:lne}

To further clarify the discussion of LNE, consider the homogenous
ferromagnet.  At positive temperature, LNE is a phenomenon separate from
the spontaneous formation of domains of the minority phase within the
majority phase.  The timescale for such a domain of size $L$ to form about
the origin is exponential in 
  (some power of) $L$.  Similarly, for a {\it finite\/} system
of size $L$, the entire system will randomly flip back and forth between
the plus and minus phases on 
  an exponential timescale.  This,
however, is not LNE, which takes place on much shorter time scales
(presumably some power of $L)$.  LNE is not due to the spontaneous
formation of one phase within another due to statistical fluctuations, but
instead is due to domain walls sweeping into the region from far away.
This contrast is even clearer at $T=0$, where the spontaneous formation of
droplets described above cannot occur; but as already discussed, in the
$2D$ ferromagnet the phenomenon of domain walls forever sweeping across any
finite region   occurs even at zero temperature.

Since the existence of LNE for all $T<T_c$ in the $2D$ Ising ferromagnet
may seem surprising, we present a possible physical mechanism which may
also shed light on LNE in general.  The initial spin configuration has
(with probability one) no infinite domains.  As the configuration evolves,
some domains shrink and others coalesce.  So the origin should always be
contained in a finite domain, whose size could usually be slowly
decreasing, but sporadically would have a large change either by coalescing
or because a domain wall passes through the origin and the identity of the
domain changes.  As a consequence we can arrive at the interesting
situation where the {\it mean\/} scale of the domain containing the origin
increases with $t$, even though, at a typical arbitrarily chosen time and
for fixed $\sigma^0$ and $\omega$, its size would be decreasing.  The
dynamical behavior averaged over initial configurations and dynamical
realizations is relatively straightforward, while for individual instances
of both it is typically complex.

Summarizing, LNE is primarily the result of nonequilibrium domain wall
motion driven by mean curvature combined with the complex domain structure
resulting from the original quench. It is also consistent with phase
separation (as would be expected from equilibrium roughening arguments).

\section{Dynamical vs.~Boltzmann averages}
\label{sec:fixed}

It was noted above that the presence of LNE might imply that some standard
assumptions regarding broken ergodicity can fail.  Suppose that LNE occurs
because only a single pair of pure states is present, such as in the $2D$
ferromagnet below $T_c$.  In such a case, would a long time average of the
magnetization in a finite region give zero?  The answer is: not
necessarily.  It could be that, after long times, the system has spent
roughly equal amounts of time in both states, in which case the usual time
average~\cite{BY86} would indeed approach zero.  But it could also happen
that, after almost any long time, the system has spent significantly more
of its life in one or the other state (which itself would change with the
observational timescale).  In other words (still using the example of a
two-state system), at any long time the weights of the two states, as
defined by a dynamical measure involving a fixed ${\omega}$ and an average
over uniformly spaced times, could be different from $1/2$, and could even
change with time (as discussed in the next section).  This is analogous to
an equilibrium phenomenon discovered by
K\"ulske~\cite{Kuelske97,Kuelske98}.  To get a zero average in this
situation one would need to average over a {\it sparse\/} sequence of
increasingly separated times.

\subsection{Chaotic Time Dependence}
\label{subsec:ctd}

We noted earlier that LNE can occur in the context of either
possibility~(2) (the averaged dynamical measure $\nu_t$ has a limit,
which is a mixed state) or (3) ($\nu_t$ does not converge). LNE {\it
must\/} occur if possibility (3) holds, but may or may not occur if
possibility~(2) holds.  We now explore further the distinctions between
these two cases.

There are two ways in which possibility~(2) can occur.  As described
earlier, for any fixed region $\Lambda_L$ the measure $\nu_{t,\tau(t)}$ can
settle into a pure state for almost all $\sigma^0$ and $\omega$, but the
pure state depends on the dynamics as well as the initial state.  But a
second possibility, that has not yet been discussed, is that
$\nu_{t,\tau(t)}$ {\it never\/} settles down to a single pure state: the
system is usually in a pure state $\alpha(t)$ locally, but the pure state
forever changes.  Nevertheless, $\nu_t$, the full average over the
dynamics, still yields a single limit.  This is to be contrasted with
possibility~(3), where even the fully averaged measure $\nu_t$ never
settles down.

Again we use the illustration of the $2D$ homogeneous ferromagnet to
clarify these statements. Below $T_c$, we know LNE occurs by Theorem~2.
Suppose furthermore that it occurs through possibility~(2).  Then, for
fixed $\sigma^0$ and at a fixed large time, for approximately half of the
dynamical realizations, a region of fixed lengthscale $L$ surrounding the
origin is in the up state (the pure state $\rho^+$), and for most of the
other half the same region is in the down state (the pure Gibbs state
$\rho^-$), and this one-to-one ratio remains essentially fixed after some
timescale depending on $L$.  Then as $t\to\infty$, $\nu_t\to\overline\rho$,
where $\overline\rho$ is the mixed Gibbs state $(1/2)\rho^+ + (1/2)\rho^-$.
Nevertheless, in any given dynamical realization, the region (as described
by $\nu_{t,\tau}$) never settles permanently into either $\rho^+$ or
$\rho^-$.

By contrast, if possibility~(3) occurs, then even the fully averaged
dynamical measure $\nu_t$ forever changes.  This could happen (again for
fixed $\sigma^0$) if the random dynamics fails to sufficiently ``mix'' the
states (in which case one has, given $\sigma^0$, some amount of predictive
power for determining from $\sigma^0$ the likely state of the system in the
region for some arbitrarily large times $t$). This is conceivable because
even though $\sigma^0$ is globally unbiased between the plus and minus
states, it does have fluctuations in favor of one or the other state of
order $\sqrt{L^2}$ on lengthscale $L$; with $L$ taken as an appropriate
power of $t(L)$, these fluctuations could (partially) predict the sign of
the phase at the origin at time $t(L)$.  In possibility~(2) on the other
hand, there is a greater capability of the random dynamics to ``mix'' the
states which eventually destroys the predictive power contained in the
fluctuations of the initial state.

So there are really two kinds of non-equilibration, corresponding either to
LNE in the framework of possibility~(2) (``weak LNE'') or else to LNE
resulting from the stronger possibility~(3).  Because $\nu_t$ evolves
deterministically according to an appropriate master equation, its lack of
a limit in possibility~(3) corresponds conceptually to the usual notion of
deterministic chaos and can thus legitimately be called chaotic time
dependence (CTD)~\cite{NS99a,FIN00}.  If weak LNE occurs, this term is not
appropriate because here the effect is due to the random dynamics.

\section{Nature vs.~Nurture}
\label{sec:nvsn}

The presence of two different ways the system can fail to equilibrate ---
weak LNE vs.~CTD --- leads to an interesting issue of predictably: if weak
LNE occurs, the configuration $\sigma^t$ for large $t$ is determined
essentially by the dynamics, and the initial configuration provides little
predictive capability in determining the state of any particular spin at a
very large time.  If CTD occurs, on the other hand, then some predictive
power from the initial configuration remains at arbitrarily large times.
This ``nature vs.~nurture'' competition provides an interesting set of
problems for future study.

Can we determine which of these possibilities occurs for selected systems?
One simple case where the extent of predictability can be precisely
determined is the $1D$ disordered ferromagnet (or spin glass) with a
continuous coupling distribution (e.g., couplings chosen uniformly from
$[0,1]$ for the ferromagnet and from the Gaussian distribution for a spin
glass).  We define (for general dimension) a dynamical order parameter
$q^t$ as~\cite{NNS00}
\begin{equation}
\label{eq:qt}
q^t = \lim_{L \to \infty} (2L+1)^{-d}\sum_{x \in \Lambda_L}
(\langle \sigma_x \rangle_t)^2 = {\bf E}_{{\cal J},\sigma^0}
(\langle \sigma_y \rangle_t^{\,2})\, .
\end{equation}
In this formula, dynamical averages (i.e., with respect to the
distribution $\nu_t$ over dynamical realizations $\omega$) of $\sigma^t$ (with
fixed ${\cal J}, \sigma^0$) are denoted by $\langle \cdot \rangle_t$
and $y$ is any fixed site, e.g., the origin; the
remaining averages, over ${\cal J}$ and $\sigma^0$, are denoted by ${\bf
E}_{{\cal J},\sigma^0}$.
The equivalence of the two formulas for $q^t$ follows from
translation-ergodicity; see~\cite{NNS00} for details.

If the infinite-time limit of $q^t$ exists, we define $q_D = \lim_{t \to
\infty}q^t$.  The order parameter $q_D$ measures the extent to which
$\sigma^\infty$ is determined by $\sigma^0$ rather than by $\omega$; it is
a dynamical analogue to the usual Edwards-Anderson order parameter.  Of
course, $q^0 = 1$ because $\sigma^0$ is completely determined by
$\sigma^0$, while a value $q_D = 0$ would mean that for every $x$, $\langle
\sigma_x \rangle_t \to 0$ so that $\sigma^0$ yields no information about
$\sigma^t$ as $t \to \infty$. The following theorem provides an exact
determination of $q_D$ for the disordered $1D$ systems introduced above:

\medskip

{\bf Theorem 3.}~\cite{NNS00} For the $d=1$ homogeneous ferromagnet at zero
temperature, $\sigma_x^\infty$ does not exist (i.e., $\sigma_x^t$ changes
infinitely many times as $t \to \infty$) for almost every $\sigma^0$ and
$\omega$ and every $x$.  For the corresponding one-dimensional disordered
model   (feromagnet or spin glass) with continuous coupling
distribution, $\sigma_x^\infty$ does exist for almost every ${\cal J}$,
$\sigma^0$, and $\omega$ and every $x$; furthermore $q_D = 1/2$.

\medskip

The value $q_D = 1/2$ is a reflection of the fact that, for almost every
${\cal J}$ and $\sigma^0$, precisely half of the $x$'s in ${\bf Z}$ have
$\sigma_x^\infty$ completely determined by $\sigma^0$ with the other
$\sigma_x^\infty$'s completely undetermined by $\sigma^0$.  For the
homogeneous one-dimensional ferromagnet, it is not hard to show (see, e.g.,
\cite{FIN00}) that although $\sigma^\infty$ does not exist, $\langle \cdot
\rangle_\infty$ does exist for almost all $\sigma^0$, and that $q_D = 0$.

What about higher-dimensional systems?  Here we mostly need to rely on
numerical studies, at least for the present.  In a recent
paper~\cite{ONSS06}, the homogeneous $2D$ ferromagnet on a square lattice
was studied at zero temperature.  Numerical results from this study suggest
that CTD might hold for the infinite lattice, in which case long-term
predictability from information contained in the initial state would be
present to some extent.  We refer the interested reader to~\cite{ONSS06} for
details.

\section{Summary}
\label{sec:summary} 

In this review, we considered the dynamical evolution of a short-range
Ising spin glass following a deep quench (although many of our results
generalize to other systems).  We presented several theorems~\cite{NS99a}
with somewhat surprising consequences, as follows. If the spin glass
displays broken spin-flip symmetry (more precisely, has a nonzero EA order
parameter), then equilibration in any local region depends crucially not
only on the number of pure states but also their overlaps.  {\it Only\/}
when there exists an {\it uncountable infinity\/} of pure states, with
almost every pair having {\it zero\/} overlap (i.e., the spin overlap
distribution $P(q)$ is a $\delta$-function at zero), can the system
equilibrate, falling into some pure state   as $t \to \infty$.
However, this is a necessary, not a sufficient condition.  

A second consequence of these theorems is that the union of the basins of
attraction of all pure states (again, if broken symmetry occurs) forms a
set of measure zero in configuration space following a deep quench: almost
every starting configuration is on a boundary between multiple pure state
basins.

This has consequences not only for deep quenches but also for slow cooling.
Once again the ferromagnet provides an instructive example.  The general
applicability of our arguments implies that the same result holds for
ferromagnets (either homogeneous or disordered) following a deep quench.
But if one cools slowly instead, then it's easy to prepare the system in
one of the two translationally invariant pure states, which are well
understood and characterized: the positive and negative magnetization
states.  

But the spin glass could present a different story under slow cooling, even
for small temperature changes. If the chaotic temperature dependence
predicted in some theories~\cite{FH88b,BM87} occurs, then the pure state
structure of a spin glass (with fixed ${\cal J}$) changes chaotically on
lengthscales larger than some $L^*(\Delta T)$ when the system undergoes a
change in temperature $\Delta T$.  The dynamical effect of such a change
may then be similar to that of a deep quench.  The well-known difficulty in
equilibrating spin glasses may therefore be a consequence of this effect,
with long relaxation times arising from small domain sizes and slow
(possibly due to pinning) motion of domain walls.

More generally, we have argued against a common viewpoint that pure state
multiplicity is irrelevant to the dynamics of infinite (or very large)
systems on finite timescales.  In many situations, a system will not spend
all of its time in a single pure state, even locally.  Because of this, it
is also not necessarily true that ``absolutely broken ergodicity'' ---
i.e., the presence of more than one pure state separated by infinite
barriers --- implies that time averages and Boltzmann averages must
disagree (or equivalently, that the limits $N\to\infty$ and $t\to\infty$
cannot commute). 

Finally, we discussed the effect of initial conditions on the future spin
configuration of a spin system, in the context of its predictability: to
what extent is the evolution determined by the starting configuration, and
how much depends on the dynamics?  This ``nature vs.~nurture'' problem can
be solved exactly for $1D$ random ferromagnets and spin
glasses~\cite{NNS00}, and was studied numerically for the $2D$ homogeneous
ferromagnet on the square lattice~\cite{ONSS06}.  The problem is equivalent
to determining whether a weak form of local non-equilibration occurs (which
favors ``nurture'') or whether a stronger chaotic time dependence occurs
(which favors ``nature'').  Which of these occurs for particular systems
remains an open problem.

\medskip

{\it Acknowledgments.\/} This work was partially supported by the National
Science Foundation under grant DMS-0604869.  We thank Joerg Rottler,
Malcolm Kennett, and Philip Stamp, guest editors of the special issue
entitled ``Classical and Quantum Glasses'' of the Journal of Physics:
Condensed Matter, for inviting us to contribute an article.

\newpage
\pagestyle{empty}

\small

\bibliographystyle{unsrt}

\bibliography{refs91007}

\small\def\em{\it} \newcommand{\noopsort}[1]{} \newcommand{\printfirst}[2]{#1}
  \newcommand{\singleletter}[1]{#1} \newcommand{\switchargs}[2]{#2#1}
\begin{thebibliography}{10}

\bibitem{BY86}
K.~Binder and A.~P. Young.
\newblock Spin glasses: experimental facts, theoretical concepts, and open
  questions.
\newblock {\em Rev. Mod. Phys.}, 58:801--976, 1986.

\bibitem{BCKM98}
P.~Bouchaud, L.~Cugliandolo, J.~Kurchan, and M.~M\'ezard.
\newblock Out of equilibrium dynamics in spin-glasses and other glassy systems.
\newblock In A.~P. Young, editor, {\em Spin Glasses and Random Fields}, pages
  161--223. World Scientific, Singapore, 1998.

\bibitem{aging1}
P.~Refrigier, E.~Vincent, J.~Hamman, and M.~Ocio.
\newblock Ageing phenomena in a spin glass: effect of temperature changes below
  ${T}_g$.
\newblock {\em J. Phys. (Paris)}, 48:1533--1539, 1987.

\bibitem{aging2}
G.~J.~M. Koper and H.~J. Hilhorst.
\newblock A domain theory for linear and nonlinear aging effects in spin
  glasses.
\newblock {\em J. Phys. (Paris)}, 49:429--443, 1988.

\bibitem{aging3}
P.~Sibani and K.-H. Hoffmann.
\newblock Hierarchical models for aging and relaxation of spin glasses.
\newblock {\em Phys. Rev. Lett.}, 63:2853--2856, 1989.

\bibitem{aging4}
K.-H. Hoffmann and P.~Sibani.
\newblock Relaxation and aging in spin glasses and other complex systems.
\newblock {\em Z. Phys. B}, 80:429--438, 1990.

\bibitem{aging5}
P.~Svedlindh, K.~Gunnarson, J.-O. Andersson, H.~A. Katori, and A.~Ito.
\newblock Time-dependent {AC} susceptibility in spin glasses.
\newblock {\em Phys. Rev. B}, 46:13867--13873, 1992.

\bibitem{aging6}
J.~P. Bouchaud.
\newblock Weak ergodicity breaking and aging in disordered systems.
\newblock {\em J. Phys. I}, 2:1705--1713, 1992.

\bibitem{aging7}
F.~Lefloch, J.~Hamann, M.~Ocio, and E.~Vincent.
\newblock Can aging phenomena discriminate between the droplet model and a
  hierarchical description in spin glasses?
\newblock {\em Europhys. Lett.}, 18:647--652, 1992.

\bibitem{aging8}
H.~Rieger.
\newblock Nonequilibrium dynamics and aging in the three-dimensional {I}sing
  spin-glass model.
\newblock {\em J. Phys. A}, 26:L615--L621, 1993.

\bibitem{aging9}
S.~Franz and M.~M\'ezard.
\newblock On mean field glassy dynamics out of equilibrium.
\newblock {\em Physica A}, 210:48--72, 1994.

\bibitem{aging10}
E.~Vincent, J.~P. Bouchaud, D.~S. Dean, and J.~Hamann.
\newblock Aging in spin glasses as a random walk: effect of a magnetic field.
\newblock {\em Phys. Rev. B}, 52:1050--1060, 1995.

\bibitem{LOHOV88}
M.~Lederman, R.~Orbach, J.~M. Hamann, M.~Ocio, and E.~Vincent.
\newblock Dynamics in spin glasses.
\newblock {\em Phys. Rev. B}, 44:7403--7412, 1988.

\bibitem{HLOOV92}
J.~M. Hammann, M.~Lederman, M.~Ocio, R.~Orbach, and E.~Vincent.
\newblock Spin-glass dynamics. {R}elation between theory and experiment: a
  beginning.
\newblock {\em Physica A}, 185:278--294, 1992.

\bibitem{JOH96}
Y.~G. Joh, R.~Orbach, and J.~Hamann.
\newblock Spin glass dynamics under a change in a magnetic field.
\newblock {\em Phys. Rev. Lett.}, 77:4648--4651, 1996.

\bibitem{VHOBC97}
E.~Vincent, J.~Hammann, M.~Ocio, J.-P. Bouchaud, and L.~Cugliandolo.
\newblock Slow dynamics and aging in spin glasses.
\newblock In M.~Rubi, editor, {\em Complex Behavior of Glassy Systems}, pages
  184--219. Springer-Verlag Lecture Notes in Physics{,} v{.} 492, Berlin, 1997.

\bibitem{JVHPN98}
K.~Jonason, E.~Vincent, J.~Hammann, J.-P. Bouchaud, and P.~Nordblad.
\newblock Memory and chaos effects in spin glasses.
\newblock {\em Phys. Rev. Lett.}, 81:3243--3246, 1998.

\bibitem{RMB00}
B.~Rinn, P.~Maass, and J.-P. Bouchaud.
\newblock Multiple scaling regimes in simple aging models.
\newblock {\em Phys. Rev. Lett.}, 84:5403--5406, 2000.

\bibitem{RTZ00}
F.~Ricc{i-T}ersenghi and R.~Zecchina.
\newblock Glassy dynamics near zero temperature.
\newblock {\em Phys. Rev. E}, 62:R7567--R7570, 2000.

\bibitem{VDAHB00}
E.~Vincent, V.~Dupuis, M.~Alba, J.~Hamann, and J.-P. Bouchaud.
\newblock Aging phenomena in spin glass and ferromagnetic phases: domain growth
  and wall dynamics.
\newblock {\em Europhys. Lett.}, 50:674--680, 2000.

\bibitem{Ogielski85}
A.~T. Ogielski.
\newblock Dynamics of thre{e-d}imensional spin glasses in thermal equilibrium.
\newblock {\em Phys. Rev. B}, 32:7384--7398, 1985.

\bibitem{OM85}
A.~T. Ogielski and I.~Morgenstern.
\newblock Critical behavior of the three-dimensional {I}sing spin-glass model.
\newblock {\em Phys. Rev. Lett.}, 54:928--931, 1985.

\bibitem{RBY90}
J.~D. Reger, R.~N. Bhatt, and A.~P. Young.
\newblock {Monte Carlo} study of the order-parameter distribution in the
  four-dimensional {I}sing spin glass.
\newblock {\em Phys. Rev. Lett.}, 64:1859--1862, 1990.

\bibitem{KY96}
N.~Kawashima and A.~P. Young.
\newblock Phase transition in the three-dimensional $\pm {J}$ {I}sing spin
  glass.
\newblock {\em Phys. Rev. B}, 53:R484--R487, 1996.

\bibitem{PY99a}
M.~Palassini and A.~P. Young.
\newblock Evidence for a trivial ground-state structure in the two-dimensional
  {I}sing spin glass.
\newblock {\em Phys. Rev. B}, 60:R9919--R9922, 1999.

\bibitem{PY99b}
M.~Palassini and A.~P. Young.
\newblock Triviality of the ground state structure in {I}sing spin glasses.
\newblock {\em Phys. Rev. Lett.}, 83:5126--5129, 1999.

\bibitem{Middleton99}
A.~A. Middleton.
\newblock Numerical investigation of the thermodynamic limit for ground states
  in models with quenched disorder.
\newblock {\em Phys. Rev. Lett.}, 83:1672--1675, 1999.

\bibitem{Middleton00}
A.~A. Middleton.
\newblock Energetics and geometry of excitations in random systems.
\newblock {\em Phys. Rev. B}, 63:060202, 2000.

\bibitem{Hartmann00}
A.~K. Hartmann.
\newblock How to evaluate ground-state landscapes of spin glasses
  thermodynamically correctly.
\newblock {\em Eur. Phys. J. B}, 13:539--545, 2000.

\bibitem{KM00}
F.~Krzakala and O.~C. Martin.
\newblock Spin and link overlaps in three-dimensional spin glasses.
\newblock {\em Phys. Rev. Lett.}, 85:3013--3016, 2000.

\bibitem{PY00}
M.~Palassini and A.~P. Young.
\newblock Nature of the spin glass state.
\newblock {\em Phys. Rev. Lett.}, 85:3017--3020, 2000.

\bibitem{HKM00}
J.~Houdayer, F.~Krzakala, and O.~C. Martin.
\newblock Large-scale low-energy excitations in 3-d spin glasses. {R}eplica
  symmetry breaking and characterization in position space.
\newblock {\em Eur. Phys. B}, 18:467--477, 2000.

\bibitem{MP00}
E.~Marinari and G.~Parisi.
\newblock Effects of changing the boundary conditions on the ground state of
  {I}sing spin glasses.
\newblock {\em Phys. Rev. B}, 62:11677--11685, 2000.

\bibitem{MP01}
E.~Marinari and G.~Parisi.
\newblock Effects of a bulk perturbation on the ground state of $3{D}$ {I}sing
  spin glasses.
\newblock {\em Phys. Rev. Lett.}, 86:3887--3890, 2001.

\bibitem{HYD04}
G.~Hed, A.~P. Young, and E.~Domany.
\newblock Lack of ultrametricity in the low-temperature phase of
  three-dimensional ising spin glasses.
\newblock {\em Phys. Rev. Lett.}, 92:157201, 2004.

\bibitem{MPV87}
M.~M\'ezard, G.~Parisi, and M.~A. Virasoro, editors.
\newblock {\em Spin Glass Theory and Beyond}.
\newblock World Scientific, Singapore, 1987.

\bibitem{MPRRZ00}
E.~Marinari, G.~Parisi, F.~Ricci-Tersenghi, J.~J. Ruiz-Lorenzo, and F.~Zuliani.
\newblock Replica symmetry breaking in spin glasses: {T}heoretical foundations
  and numerical evidences.
\newblock {\em J. Stat. Phys.}, 98:973--1047, 2000.

\bibitem{Mac84}
W.~L. McMillan.
\newblock Scaling theory of {I}sing spin glasses.
\newblock {\em J. Phys. C}, 17:3179--3187, 1984.

\bibitem{BM85}
A.~J. Bray and M.~A. Moore.
\newblock Critical behavior of the three-dimensional {I}sing spin glass.
\newblock {\em Phys. Rev. B}, 31:631--633, 1985.

\bibitem{FH86}
D.~S. Fisher and D.~A. Huse.
\newblock Ordered phase of short-range {I}sing spin-glasses.
\newblock {\em Phys. Rev. Lett.}, 56:1601--1604, 1986.

\bibitem{FH88a}
D.~S. Fisher and D.~A. Huse.
\newblock Nonequilibrium dynamics of spin glasses.
\newblock {\em Phys. Rev. B}, 38:373--385, 1988.

\bibitem{FH88b}
D.~S. Fisher and D.~A. Huse.
\newblock Equilibrium behavior of the spin-glass ordered phase.
\newblock {\em Phys. Rev. B}, 38:386--411, 1988.

\bibitem{BM87}
A.~J. Bray and M.~A. Moore.
\newblock Chaotic nature of the spin-glass phase.
\newblock {\em Phys. Rev. Lett.}, 58:57--60, 1987.

\bibitem{NS97}
C.~M. Newman and D.~L. Stein.
\newblock Metastate approach to thermodynamic chaos.
\newblock {\em Phys. Rev. E}, 55:5194--5211, 1997.

\bibitem{NS98}
C.~M. Newman and D.~L. Stein.
\newblock Simplicity of state and overlap structure in finite-volume realistic
  spin glasses.
\newblock {\em Phys. Rev. E}, 57:1356--1366, 1998.

\bibitem{NS03b}
C.~M. Newman and D.~L. Stein.
\newblock Topical {R}eview: {O}rdering and broken symmetry in short-ranged spin
  glasses.
\newblock {\em J. Phys.: Cond. Mat.}, 15:R1319 -- R1364, 2003.

\bibitem{NS99c}
C.~M. Newman and D.~L. Stein.
\newblock Metastable states in spin glasses and disordered ferromagnets.
\newblock {\em Phys. Rev. E}, 60:5244--5260, 1999.

\bibitem{EA75}
S.~Edwards and P.~W. Anderson.
\newblock Theory of spin glasses.
\newblock {\em J. Phys. F}, 5:965--974, 1975.

\bibitem{Bray94}
A.~J. Bray.
\newblock Theory of phase-ordering kinetics.
\newblock {\em Adv. Phys.}, 43:357--459, 1994.

\bibitem{NS06a}
C.~M. Newman and D.~L. Stein.
\newblock Local vs. global variables for spin glasses.
\newblock In E.~Bolthausen and A.~Bovier, editors, {\em Spin Glass Theory},
  pages 145--158. Springer, Berlin, 2006.

\bibitem{NS99a}
C.~M. Newman and D.~L. Stein.
\newblock Equilibrium pure states and nonequilibrium chaos.
\newblock {\em J. Stat. Phys.}, 94:709--722, 1999.

\bibitem{NNS00}
S.~Nanda, C.~M. Newman, and D.~L. Stein.
\newblock Dynamics of {I}sing spin systems at zero temperature.
\newblock In R.~Minlos, S.~Shlosman, and Y.~Suhov, editors, {\em On Dobrushin's
  Way (from Probability Theory to Statistical Physics)}, pages 183--194. Amer.
  Math. Soc. Transl. (2) 198, 2000.

\bibitem{Stauffer94}
D.~Stauffer.
\newblock Ising spinodal decomposition at ${T=0}$ in one to five dimensions.
\newblock {\em J. Phys. A}, 27:5029--5032, 1994.

\bibitem{vEvH84}
A.~C.~D. {v}an Enter and J.~L. {v}an Hemmen.
\newblock Statistical-mechanical formalism for spin-glasses.
\newblock {\em Phys. Rev. A}, 29:355--365, 1984.

\bibitem{Bag96}
F.~Bagnoli.
\newblock On damage-spreading transitions.
\newblock {\em J. Stat. Phys.}, 85:151--164, 1996.

\bibitem{Grass95}
P.~Grassberger.
\newblock Are damage spreading transitions generically in the universality
  class of directed percolation?
\newblock {\em J. Stat. Phys.}, 79:13--23, 1995.

\bibitem{JR94}
N.~Jan and T.~S. Ray.
\newblock {``Damage''} in the low-temperature phase of the $\pm {J}$ spin glass
  in two to six dimensions.
\newblock {\em J. Stat. Phys.}, 75:1197--1204, 1994.

\bibitem{KH88}
G.~J.~M. Koper and H.~J. Hilhorst.
\newblock A domain theory for linear and nonlinear aging effects in spin
  glasses.
\newblock {\em J. Phys. (Paris)}, 49:429--443, 1988.

\bibitem{TH96}
M.~J. Thill and H.~J. Hilhorst.
\newblock Theory of the critical state of lo{w-d}imensional spin glass.
\newblock {\em J. Phys. I}, 6:67--95, 1996.

\bibitem{Palmer82}
R.~G. Palmer.
\newblock Broken ergodicity.
\newblock {\em Adv. Phys.}, 31:669--735, 1982.

\bibitem{GNS00}
A.~Gandolfi, C.M. Newman, and D.L. Stein.
\newblock Zero temperature dynamics of $\pm {J}$ spin glasses and related
  {I}sing models.
\newblock {\em Commun. Math. Phys.}, 214:373--387, 2000.

\bibitem{Kuelske97}
C.~K{\"u}lske.
\newblock Limiting behavior in random {G}ibbs measures: {M}etastates in some
  disordered mean field models.
\newblock In A.~Bovier and P.~Picco, editors, {\em Mathematics of Spin Glasses
  and Neural Networks}, pages 151--160. Birkhauser, Boston, 1998.

\bibitem{Kuelske98}
C.~K{\"u}lske.
\newblock Metastates in disordered mean-field models {II}: The superstates.
\newblock {\em J. Stat. Phys.}, 91:155--176, 1998.

\bibitem{FIN00}
L.~R.~G. Fontes, M.~Isopi, and C.~M. Newman.
\newblock Random walks with strongly inhomogeneous rates and singular
  diffusions: convergenc{e,} localization and aging in one dimension.
\newblock {\em Ann. Probab.}, 30:579--604, 2002.

\bibitem{ONSS06}
P.~M.~C. {de O}liveira, C.~M. Newman, V.~Sidoravicious, and D.~L. Stein.
\newblock Ising ferromagnet: Zero-temperature dynamical evolution.
\newblock {\em J. Phys. A}, 39:6841--6849, 2006.

\end{thebibliography}

\end{document}